\documentclass{article}
\usepackage{ltwol2e}
\font\bfit=cmmib10

\def \ts{\thinspace}
\def \Wg{W\negthinspace g}

\def \GeV{{\rm \enspace GeV}}
\def \TeV{{\rm \enspace TeV}}
\def \beq{\begin{equation}}
\def \eeq{\end{equation}}
\def \beqa{\begin{eqnarray}}
\def \eeqa{\end{eqnarray}}
\def \Aud{{\cal A_{\uparrow\downarrow}}}
 
\begin{document}


\title{Spin Polarization in Single Top Events}

\author{Gregory Mahlon$^{*}$}

\address{Department of Physics, McGill University, 
3600 University St., Montr\'eal, QC  H3A 2T8, Canada\\
Electronic address:  mahlon@physics.mcgill.ca}

\date{October 29, 1998}

\twocolumn[\maketitle\abstracts{
We discuss the optimal spin bases for describing angular
correlations in single top quark events at the Fermilab Tevatron.
We define spin bases that exploit the fact that  
the top quarks are produced with 100\% polarization
along the momenta of the $d$-type quarks in these events.
For single top production through an $s$-channel $W$ boson, 
98\% of the top quarks have their spins in the antiproton direction.
For single top production via $\Wg$-fusion, 96\% of the
top quarks have their spins in the spectator jet direction.
The direction of 
the top quark spin is reflected in the angular correlations
of its decay products.
\hfill McGill/98-32
}]


Until the discovery of the top quark, most studies of spin
in high energy physics were formulated in terms of the
helicity basis.  For ultrarelativistic particles, this 
is appropriate.  However, in general, the direction and degree
of polarization of a massive spinning particle depends on
how it was produced.  Thus, for moderate particle energies,
it should not be surprising to find that the optimal axis
for studying spin correlations is something other than 
the particle's direction of motion.

For the case of single top quark production at the Tevatron,
all of the Standard Model diagrams contain a common subdiagram:
somewhere there is a $ud$ quark line which is attached to
a $tb$ quark line via a $W$ boson.  The exact orientation
of this subdiagram depends on the process being considered.
However, the fact that the $W$ boson couples only to 
fermions with left-handed chirality
leads to a 100\% polarization of the produced top quark
in the direction of the $d$-type quark.\cite{OptimalBasis}
Thus, in studying the spin correlations in single top production
at the Tevatron the two key questions are ``Where is the
$d$-type quark?'' and ``How well can we know the location
of the $d$-type quark?''


Before answering these questions, let us clarify exactly what we
mean when we state that the top quark spin points in the direction
of the $d$-type quark.
As the first step, consider the
correlations among the decay products of 
a spin up top quark.  The dominant Standard Model decay
chain is
\beq
\includegraphics{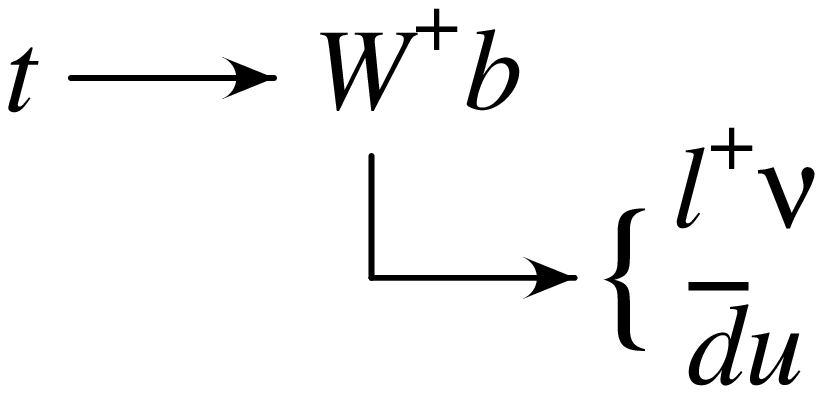}
\phantom{\Bigl[\over\Bigl[}
\eeq
For concreteness, we will describe the leptonic $W$ decay.
However, everything which we say about the charged lepton
applies equally to the $d$-type quark in a hadronic decay.

\begin{figure}[t]
\vskip4.2cm
\includegraphics{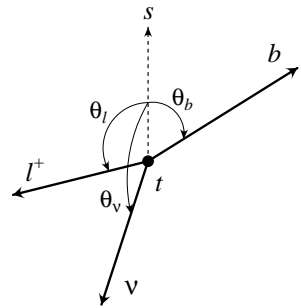}
\caption[]{Definition of the top quark decay angles in the
top quark rest frame.  The direction of the top quark spin
is indicated by the vector $s$.  Although we have drawn this
figure assuming a leptonic $W$ decay, the same correlations
hold in a hadronic decay if we replace the charged lepton by
the $d$-type quark and the neutral lepton by the $u$-type
quark.}
\label{DecayAngles}
\end{figure}

\begin{figure}[t]
\vskip4.5cm
\includegraphics{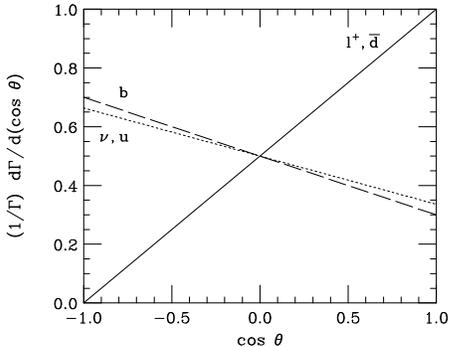}
\caption[]{Angular correlations in the decay of a spin up top
quark.  The lines labeled $\ell^{+}$, $\bar{d}$, $b$, $\nu$,
and $u$ are the angle between the spin axis and the particle
in the rest frame of the top quark.}
\label{dGammaPlot}
\end{figure}

We define the decay angles in the top quark rest frame
with respect to top quark spin vector $s$,
as shown in Fig.~\ref{DecayAngles}.
The decay angular distributions are simply linear
in the cosine of these decay angles:
\beq
{1\over\Gamma}\thinspace
{ {d\Gamma}\over{d(\cos\theta_i)} }
=
{1\over2}
\Bigl( 1+\alpha_i\cos\theta_i \Bigr),
\label{dGamma}
\eeq
where $\theta_i$ is the decay 
angle of the $i$th decay product.\cite{alphas}
The degree to which each decay product is correlated
with the spin is encoded in the value of $\alpha_i$.
From the listing of $\alpha_i$ values in Table~\ref{alphaTable},
we see that
the charged lepton is maximally correlated.
Thus, the most distinctive distribution plots the angle between
the spin axis and
the charged lepton in the top quark rest frame 
(see Fig.~\ref{dGammaPlot}).

\begin{table}[t]
\centering%
\caption{Correlation coefficients $\alpha_i$ for both
semileptonic and hadronic top quark decays.  The first
two entries are a function of $M_t^2/M_W^2$, and have
been evaluated for  
$M_t = 173.8 \GeV$ and $M_W = 80.41 \GeV$.\protect\cite{PDG,fns}
\lower2pt\hbox{\protect\phantom{j}}
\label{alphaTable}}
\begin{tabular}{ccccc}
\multispan5\hrulefill \\[0.05cm]
&Decay Product &\quad& $\alpha_i$ & \\[0.2cm]
\multispan5\hrulefill \\[0.1cm]
\qquad\qquad\quad\thinspace\enspace
& $b$  &\quad&  $-0.40$  & \\
& $\nu_{\ell}, u,$ or $c$
     &\quad& $-0.33$ & \thinspace\enspace\quad\qquad\qquad \\
& $\bar{\ell}, \bar{d},$ or $\bar{s}$
     &\quad& $\protect\phantom{-}1.00$ & \qquad\quad \\[0.1cm]
\multispan5\hrulefill 
\end{tabular}
\end{table}

When we write the decay matrix element in an arbitrary
Lorentz frame, we find that
the natural 4-vectors are not
the top quark momentum $t$ and its spin 
vector $s$ (normalized such that $s_\mu s^\mu = -1$).
Instead, it is more convenient to use the combinations
\beq
t_1 \equiv \hbox{$1\over2$} (t+Ms) \quad{\rm and}\quad
t_2 \equiv \hbox{$1\over2$} (t-Ms),
\eeq
where $M$ is the mass of the top quark.  
In the top quark rest frame, the spatial parts of $t_1$ and $s$
point in the same direction, since in this frame $t = (M,{\bf 0})$.
In some other frame, however, these vectors
are not parallel.  In this case, the form of the
matrix element clearly indicates
that the preferred charged lepton emission axis is the
spatial part of $t_1$.
Hence, we should regard $t_1$ as the
appropriate generalization of the spin axis to an arbitrary
reference frame.  

Finally, we observe that $t_1$ is a massless vector.
Thus, we may specify it by using the 4-momentum of any massless
particle in the process being studied.  In particular, for
single top production, we find that the spin down polarized
production amplitude contains a factor of $d\cdot t_1$,
where $d$ is the 4-momentum of the $d$-type quark.
Choosing $t_1 = d$ causes this amplitude to vanish, 
meaning that the entire production cross section comes
from the spin up amplitude.
This is what we mean by saying that
the top quarks are 100\% polarized in the
direction of the $d$-type quark.  Note that the preferred
emission direction 
of the charged lepton from the decaying top quark
as viewed {\it in any frame}
is the same as the direction of the $d$-type quark 
in that frame.

As we shall see shortly, it is not always possible to
know the location of the $d$-type quark.  The
best we can do is choose a spin axis that maximizes
the likelihood that we have picked the ``correct'' direction.
In this situation, we will have a mixed sample containing
$N_\uparrow$ spin-up and $N_\downarrow$
spin-down top quarks, and the    
angular distribution in Eq.~(\ref{dGamma})
must be replaced by
\beq
{1\over\Gamma}\thinspace
{ {d\Gamma}\over{d(\cos\theta_{i})} }
=
{1\over2}
\biggl(1+
\Aud \ts
\alpha_\ell \ts 
\cos\theta_{i}\biggr),
\label{dGammaMixed}
\eeq
where $\Aud$ is the spin asymmetry
\beq
\Aud \equiv
{ {N_\uparrow-N_\downarrow}
\over
{N_\uparrow+N_\downarrow} }.
\label{SpinAsym}
\eeq
Eq.~(\ref{dGammaMixed}) tells us that we obtain the 
largest correlations
by using the charged lepton and 
choosing the spin axis which maximizes the magnitude of
$\Aud$ -- {\it i.e.}\ the best approximation to the 
direction of the $d$-type quark.

This brings us to the two key questions proposed
at the beginning of this talk:
``Where is the $d$-type quark?''
and ``How well can we know the location of the $d$-type quark?''


\begin{figure}[t]
\vskip2.7cm
\includegraphics{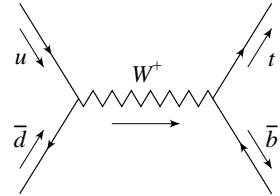}
\caption[]{Feynman diagram for single top quark production in the
$W^{*}$ process.}
\label{WstarDiagram}
\end{figure}

We begin with a discussion of the $W^{*}$ production 
mechanism,\cite{WstarRefs}
also known as the $s$-channel production mechanism, because
the off-shell $W$ is in the $s$-channel
(see Fig.~\ref{WstarDiagram}).  Conceptually, this
is a very simple process:  a $u$ quark and a $\bar{d}$ quark
annihilate, forming  an off-shell $W$ which ``decays'' to a $t$ quark
and a $\bar{b}$ quark.  
Since the proton beam
is a copious source of quarks and the antiproton beam a
copious source of antiquarks,
we intuitively expect that the $d$-type quark
will come from the antiproton beam most of the time.
In fact, we find that the antiproton beam supplies the $d$-type
quark 98\% of the time at a center of mass energy $\sqrt{s} = 2.0 \TeV$
(see Table~\ref{WstarProd}).
Thus, we define the {\it antiproton basis}\ to be that 
basis where the top quark spin is measured along the di-

\begin{table}[h]
\centering
\caption{Fractional cross sections for single top quark
production in the $W^{*}$ channel at the Tevatron 
with $\protect\sqrt{s}= 2.0 \TeV$,
decomposed according to the parton content of the initial
state.\protect\cite{PDFs}
\lower2pt\hbox{\protect\phantom{j}}
\label{WstarProd}}

\begin{tabular}{cc@{\qquad}c@{\qquad}cc}
\multispan5\hrulefill \\[0.05cm]
\qquad\qquad\qquad\qquad
&   $p$     & $\bar{p}$ & Fraction & \qquad\qquad\qquad\qquad \\[0.1cm]
\multispan5\hrulefill \\[0.1cm]
&   $u$     & $\bar{d}$ & 98\% & \\
& $\bar{d}$ &   $u$     & \protect\phantom{0}2\% & \\[0.1cm]
\multispan5\hrulefill 
\end{tabular}
%
%
\caption{Dominant spin fractions and asymmetries for the helicity
and antiproton bases for single top quark production in the $W^{*}$
channel at the Tevatron with $\protect\sqrt{s} = 2.0 \TeV$.
\lower2pt\hbox{\protect\phantom{j}}
\label{WstarFractions}}

\begin{tabular}{ccccc}
\multispan5\hrulefill \\[0.05cm]
&Basis      & Spin Content & $\Aud$ &  \\[0.1cm]
\multispan5\hrulefill \\[0.1cm]
\ts\ts\enspace\enspace\quad\qquad
&helicity   &   83\% L          & $-0.66$ &\qquad\quad\ts\ts \\
&antiproton &   98\% $\Uparrow$ & $+0.96$ & \\[0.1cm]
\multispan5\hrulefill 
\end{tabular}
\end{table}

\noindent 
rection of the antiproton beam momentum.  We compare 
this basis to the helicity basis in Table~\ref{WstarFractions}.
Notice that in the antiproton basis, the spin
asymmetry is $\Aud = 0.96$,
whereas in the helicity basis $\Aud$ is 
only $-0.66$.\footnote{We equate spin-up with right-handed
helicity and spin-down with left-handed helicity.}
Thus, the correlations are a factor of 1.45 larger in the
antiproton basis than in the helicity basis.


\begin{figure}[t]
\vskip5.3cm
\includegraphics{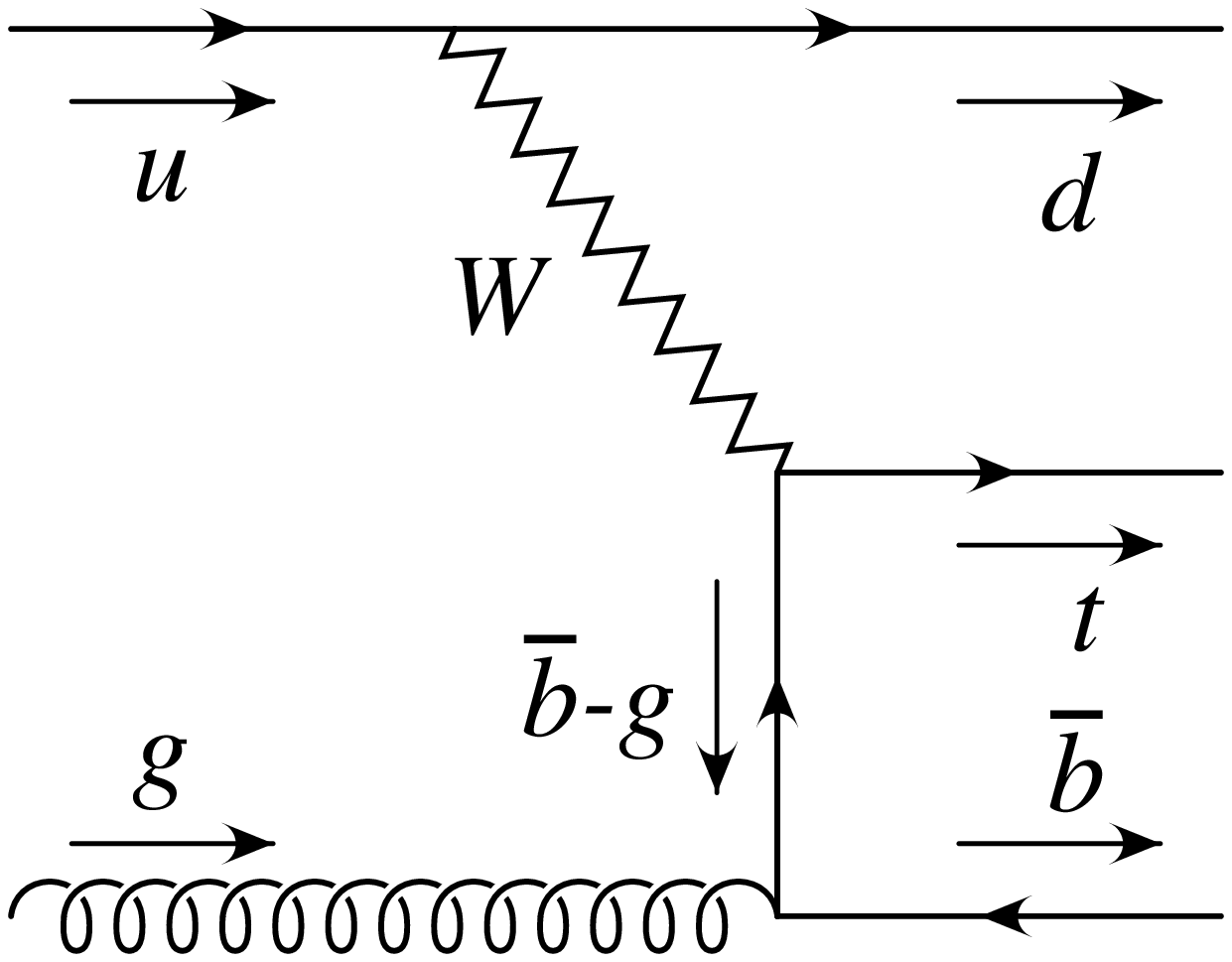}
\includegraphics{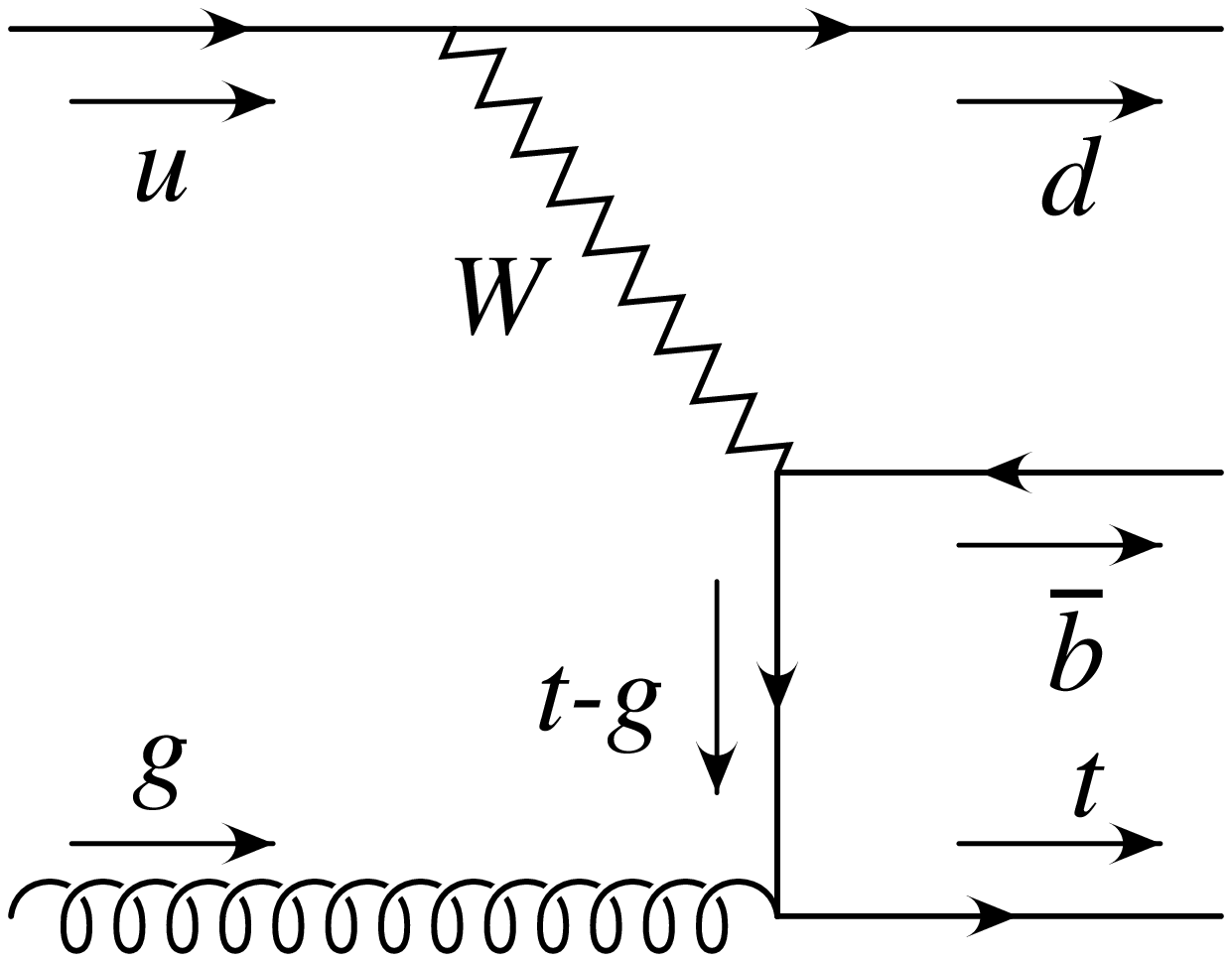}
\includegraphics{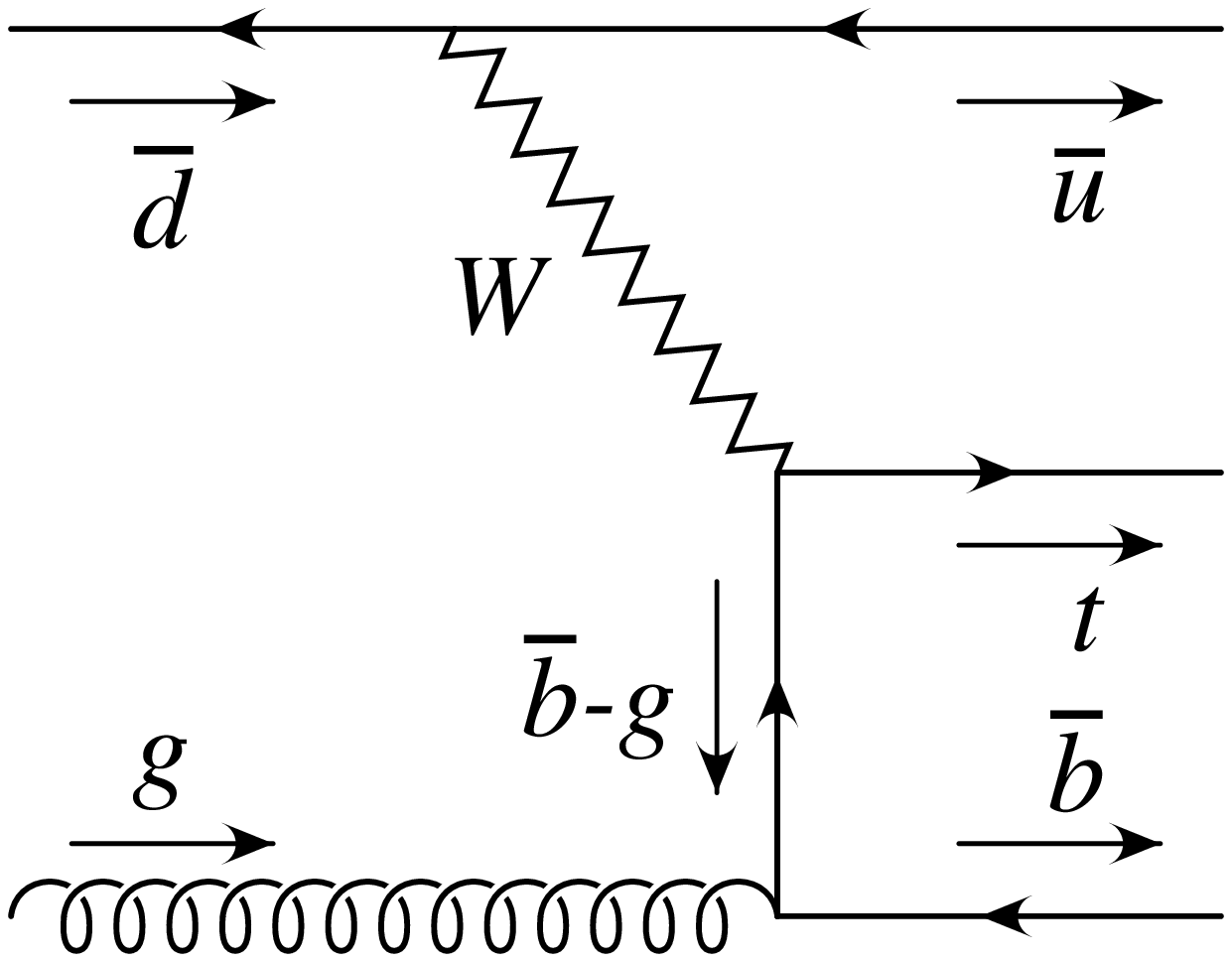}
\includegraphics{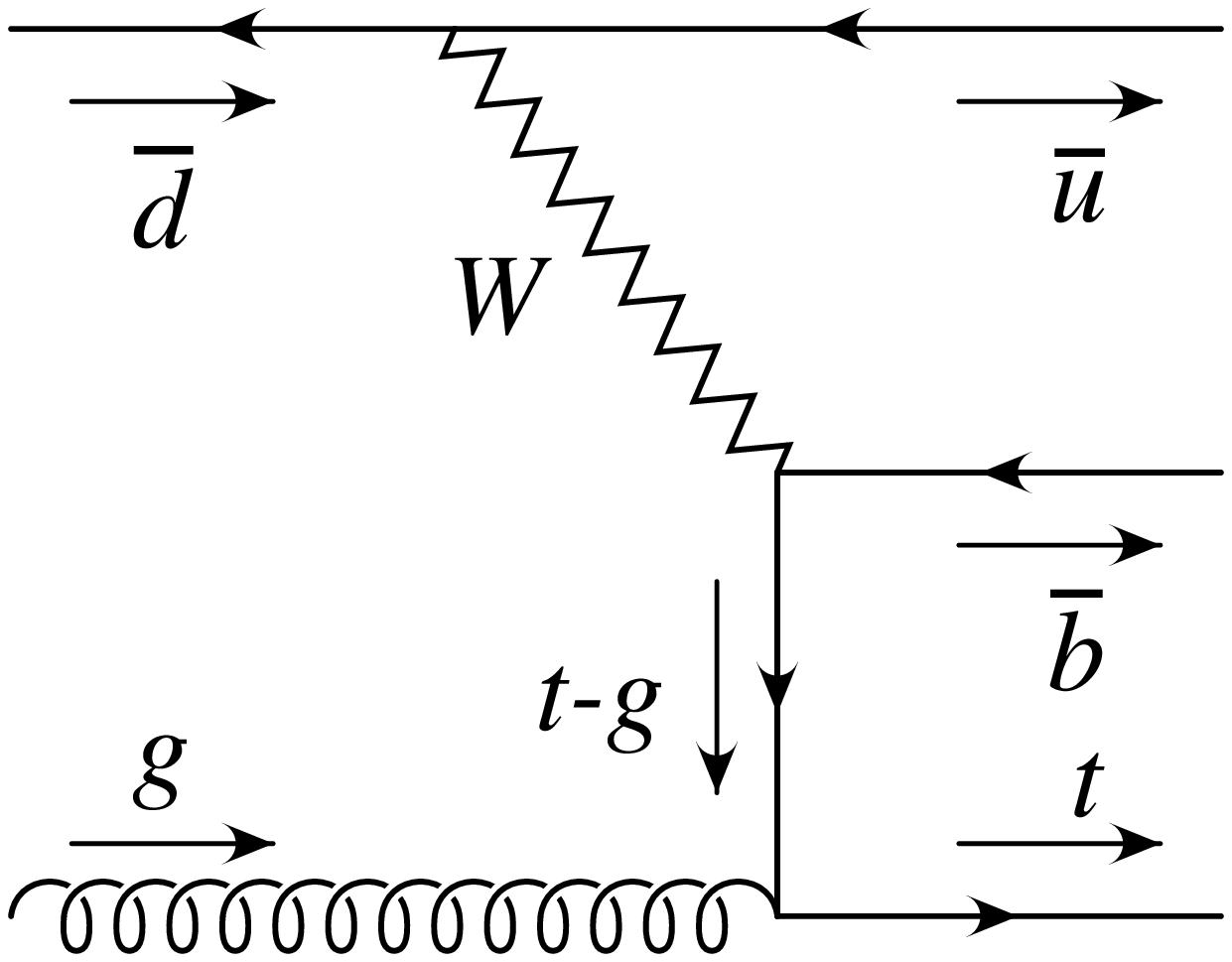}
\caption[]{Gauge-invariant sets of Feynman diagrams for single
top quark production via $\Wg$-fusion.  The lower two diagrams
are related to the upper two diagrams by crossing symmetry.}
\label{WgDiagram}
\end{figure}

The other, and, in fact, dominant production mechanism for 
single top quarks
at the Tevatron at 2.0 TeV is the so-called $W$-gluon fusion
process.\cite{WgRefs,Justification}
For the purposes of determining the spin correlations
of these events at lowest order, it is sufficient to consider
the two pairs of diagrams in
Fig.~\ref{WgDiagram}.\cite{OptimalBasis,Justification}
In these
diagrams, a gluon from one beam
splits into a $b\bar{b}$ or $t\bar{t}$ pair,
which fuses with a $W$ radiated from a light ($u$ or $\bar{d}$\ts)
spectator quark from the other beam. 
The final state contains a $\bar{b}$ jet, the top quark
decay products, and the spectator jet.

Since the $d$-type quark is either contained in one of the
beams or in the spectator jet, we should choose the 
spin axis direction
(proton momentum, antiproton momentum, or spectator jet momentum)
which maximizes the probability that our spin axis is
aligned with the $d$-type quark.  
Now, the $u$ quark content  of the proton is greater
than the $\bar{d}$ quark content of the antiproton.
Furthermore, the gluon content of both is the same.
Hence, we expect that the largest share of the cross
section comes from $ug \rightarrow t\bar{b}d$, with
the spectator jet containing the $d$ quark.  As we can
see from Table~\ref{WgProd}, this expectation is correct.
In fact, the spectator jet is the $d$ quark 77\% of the time.
Thus, we define the {\it spectator basis}\ as the basis
in which we choose the spin axis to be aligned with the
momentum of the spectator jet.  In this basis, the top
quark is produced in the spin up state 96\% of the time,
leading to correlations which are a factor of 1.35 larger
than in the helicity basis (see Table~\ref{WgFractions}).

\begin{table}
\centering
\caption{Fractional cross sections for single top quark
production in the $\Wg$-fusion channel at the Tevatron 
with $\protect\sqrt{s} = 2.0 \TeV$,
decomposed according to the parton content of the initial
state.\protect\cite{PDFs}
\lower2pt\hbox{\protect\phantom{j}}
\label{WgProd}}

\begin{tabular}{cc@{\qquad}c@{\qquad}cc}
\multispan5\hrulefill \\[0.05cm]
\qquad\qquad\qquad\qquad
&   $p$     & $\bar{p}$ & Fraction & \qquad\qquad\qquad\qquad \\[0.1cm]
\multispan5\hrulefill \\[0.1cm]
&   $u$     &   $g$     & 74\% & \\
&   $g$     &   $u$     & \protect\phantom{0}3\% & \\
&   $g$     & $\bar{d}$ & 20\% & \\
& $\bar{d}$ &   $g$     & \protect\phantom{0}3\% & \\[0.1cm]
\multispan5\hrulefill 
\end{tabular}
\end{table}

\begin{table}
\centering
\caption{Dominant spin fractions and asymmetries for the helicity
and spectator bases for single top quark production in the $\Wg$-fusion
channel at the Tevatron 
with $\protect\sqrt{s} =  2.0 \TeV$.
\lower2pt\hbox{\protect\phantom{j}}
\label{WgFractions}}

\begin{tabular}{ccccc}
\multispan5\hrulefill \\[0.05cm]
&Basis      & Spin Content & $\Aud$ & \\[0.1cm]
\multispan5\hrulefill \\[0.1cm]
\qquad\quad\ts
&helicity   &   83\% L          & $-0.67$ &\qquad\quad\ts \\
&spectator  &   96\% $\Uparrow$ & $+0.91$ & \\[0.1cm]
\multispan5\hrulefill 
\end{tabular}
\end{table}

\begin{figure}[t]
\vskip4.9cm
\includegraphics{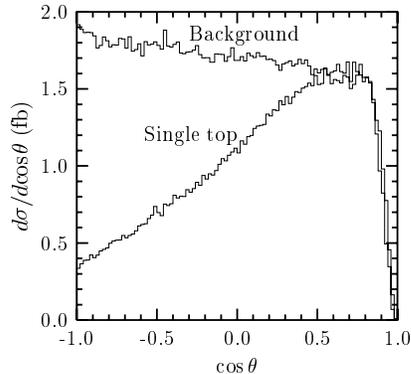}
\caption[]{Monte Carlo results of 
Stelzer, {\it et al.}\ts\protect\cite{Study}, for the
top quark rest frame
angular distribution of the charged lepton in single
top quark events at the Tevatron ($\protect\sqrt{s}=2 \TeV$),
with respect to the untagged jet. 
Also shown is the angular distribution of the background
events passing their selection criteria.  For details, see
Ref.~\protect\ref{Study}.}
\label{StelzerFigure}
\end{figure}

We now briefly address the question of actually trying to
observe these correlations in a collider environment
such as the Tevatron.
Stelzer {\it et al.}\ts\cite{Study} have simulated the $\Wg$-fusion
process at 2 TeV including backgrounds and the effects of
energy smearing, jet reconstruction, and cuts
(for details, see Ref.~\ref{Study}).
In their examination of spin correlations, they consider
a sample of events where there is exactly one tagged and
one untagged jet with transverse momentum above 20 GeV
(to control the background from $t\bar{t}$\ts).  
They assume
that the untagged jet is the spectator jet, and construct
the lepton decay angle in the top quark rest frame,
which has been determined by choosing the solution for
the neutrino momentum with the smallest magnitude of rapidity.
Their results are shown in Fig.~\ref{StelzerFigure}.  
The bins near $\cos\theta = 1$ are depleted by the
isolation cut imposed on the lepton to distinguish it from
the spectator jet.  Away from this region, the signal
shows a distinct slope, while the background is nearly flat.

To quantify the size of the correlations
present in Fig.~\ref{StelzerFigure},
Stelzer {\it et al.}\ 
define a
cross section asymmetry
over the range $-1 \leq \cos\theta \leq 0.8$, excluding
the region where the cuts are most troublesome.
Specifically,
\beq
A \equiv
{
{ \sigma(-1{\le}\cos\theta{\le}-0.1) 
  - \sigma(-0.1{\le}\cos\theta{\le}0.8) }
\over
{ \sigma(-1{\le}\cos\theta{\le}-0.1) 
  + \sigma(-0.1{\le}\cos\theta{\le}0.8) }
}.
\eeq
If we assume that we correctly identify the spectator jet 100\%
of the time and measure all quantities perfectly (the ideal
case), then $\Aud = 0.91$ and Eq.~(\ref{dGammaMixed})
tells us that $A=-45\%$.
The authors of Ref.~\ref{Study} report that for the signal
distribution in Fig.~\ref{StelzerFigure}, $A=-38\%$.  The difference
is attributable to the effects of smearing, cuts, 
the neutrino momentum reconstruction uncertainty,
misidentification of the spectator jet, etc.  
Of course, what would
be measured in a real experiment is the sum of the signal and
background distributions, in which case $A$ is further reduced
to $-14\%$.  Nevertheless, with the 2 fb$^{-1}$ of integrated
luminosity expected at Run II, this asymmetry should be visible at
the $3\sigma$ level.  To reach the $5\sigma$ level, 
a total of 5 fb$^{-1}$ is required.

In conclusion, we have found that the natural spin axis in
single top events is the direction of the $d$-type quark:
the top quarks are 100\% polarized in this direction.\cite{OptimalBasis}
Since the exact direction of the $d$-type quark is
unknown, it is necessary to choose the direction which
is most likely to be correct.  In particular, for 
production through an $s$-channel $W$ boson,
we define the antiproton basis, where the spin axis is
the direction of the antiproton beam, since this is the location
of the $d$-type quark 98\% of the time.  We find that
the correlations are a factor of 1.45 larger in the antiproton
basis than in the helicity basis.  For production via
$\Wg$-fusion, we define the spectator basis, where the
spin axis is the direction of the spectator jet.  This
is correct 77\% of the time, and leads to correlations which
are a factor of 1.35 larger than in the helicity basis.
A recent study\ts\cite{Study}
has shown for the $\Wg$-fusion mode that these correlations
will be observable at the $3\sigma$ level in Tevatron Run II.


\section*{APPENDIX: {\bfit W}\ Boson Spin in Top Quark Decay}

\begin{figure}[t]
\vskip4.2cm

\includegraphics{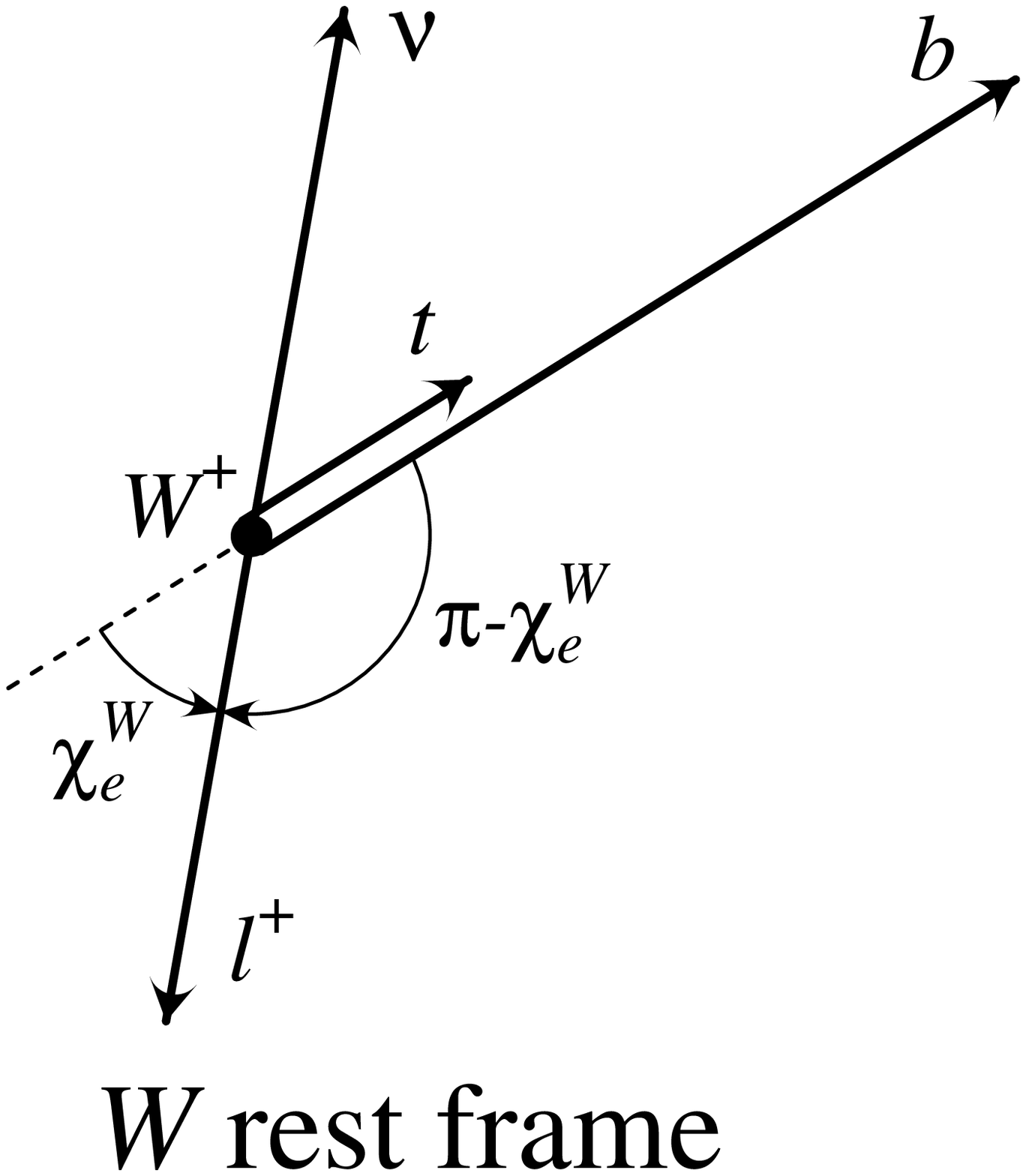}
\includegraphics{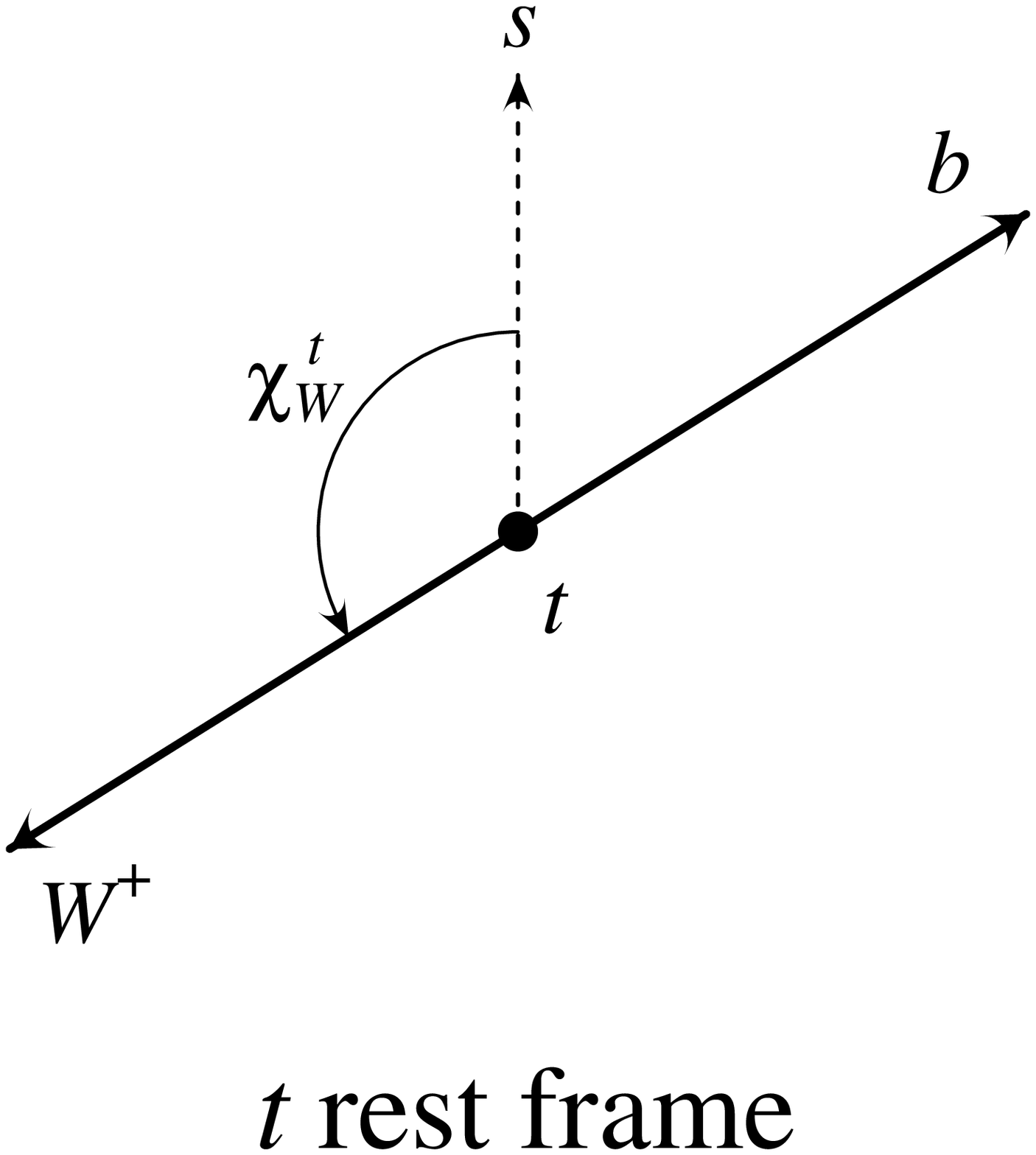}
\caption[]{Definition of the $W$ decay angles in top quark decay.
We define $\pi-\chi_e^{W}$ to be the angle between the $b$
quark direction and the charged lepton direction in the $W$
rest frame.  The other interesting angle is $\chi_W^{t}$,
the angle between the $W$ boson momentum and the top quark
spin axis in the top quark rest frame.
}
\label{Wdecay}
\end{figure}

\begin{figure}[t]
\vskip5.7cm
\includegraphics{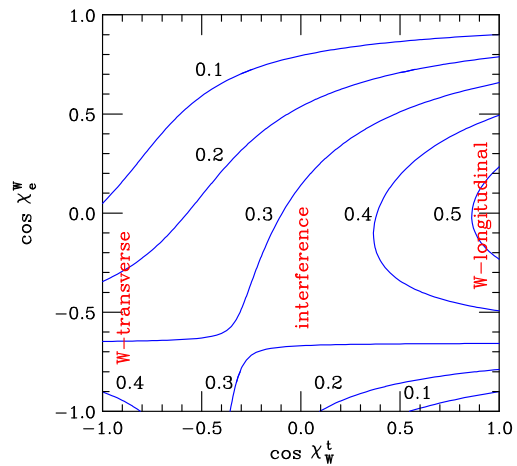}
\caption[]{Contours of the top quark decay distribution
in the $\cos\chi_W^t$-$\cos\chi_e^W$ plane.\protect\cite{Winf}
$W$ bosons
emitted in the forward direction ($\cos\chi_W^t\sim 1$)
are primarily longitudinal, while backward-emitted $W$'s
are mostly transverse.  Interference between the two spin
orientations dominates in the region around $\cos\chi_W^t=0$.}
\label{W2D}
\end{figure}

It is well-known that the $W$ bosons emitted from decaying
top quarks have a specific 
mixture of helicities\ts\footnote{It turns out that the
helicity basis is the only basis in which only two of the
three spin states contribute.  Hence, it is the most natural
basis for discussing the $W$ spin in this setting.}
in the Standard Model:  the right-handed helicity state
is absent, while the left-handed and longitudinal states
are present in the ratio $2M_W^2 : M_t^2$.  This
is reflected in the charged lepton decay distribution in the
$W$ rest frame
\beq
{1 \over \Gamma}
{{d\Gamma}
\over
{d(\cos\chi_e^{W})}
}
= { 3\over4 } \thinspace
{ {\thinspace m_t^2 \sin^2 \chi_e^{W} + 
 2m_W^2 \hbox{$1\over2$}(1-\cos \chi_e^{W})^2}
\over
{m_t^2 + 2m_W^2} },
\eeq
where the angle $\chi_e^{W}$ is defined in Fig.~\ref{Wdecay}.
Parke and Shadmi\ts\cite{Winf} have pointed out an interesting 
correlation
between this angle, $\chi_e^{W}$, and the direction that the
$W$ was emitted with respect to the top quark spin axis
in the top quark rest frame ({\it cf.}\ $\chi_W^t$,
see Fig.~\ref{Wdecay}).
This correlation, which is caused by the interference between
the two (unobserved) spin states of the $W$ boson, is illustrated
in Fig.~\ref{W2D}.  It is apparent from this distribution that
as viewed in the top quark rest frame,
the longitudinal $W$ bosons  are emitted preferentially in the
same direction as the top quark spin, whereas the transverse
$W$'s prefer the backwards direction relative to the spin axis.

\begin{figure}[t]
\vskip4.6cm
\includegraphics{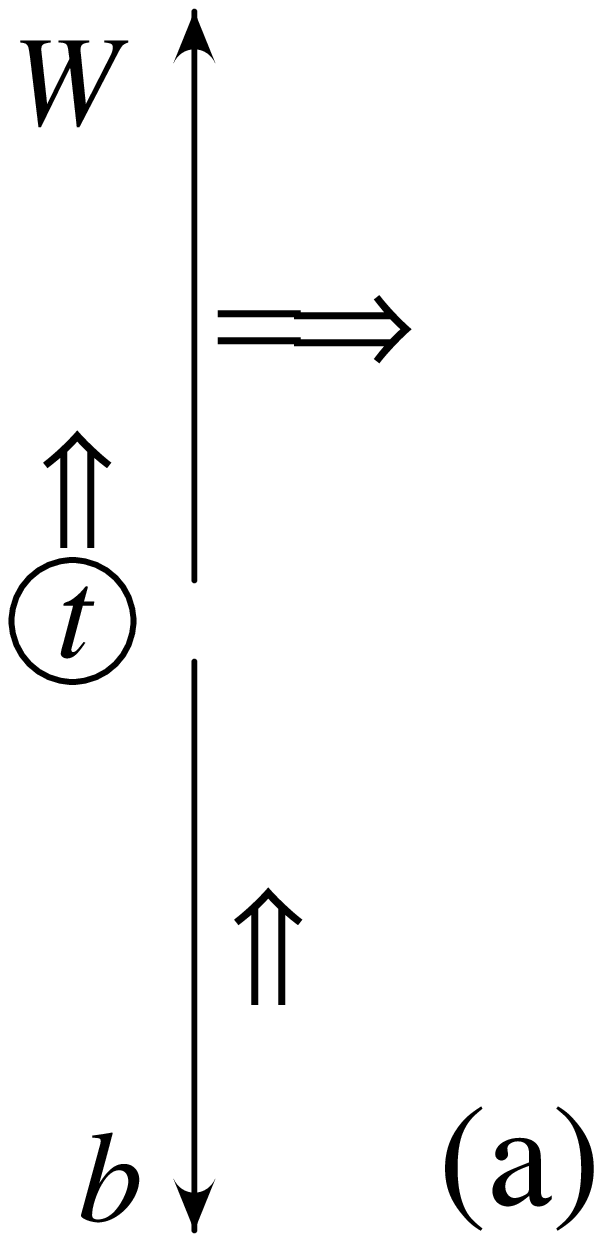}
\includegraphics{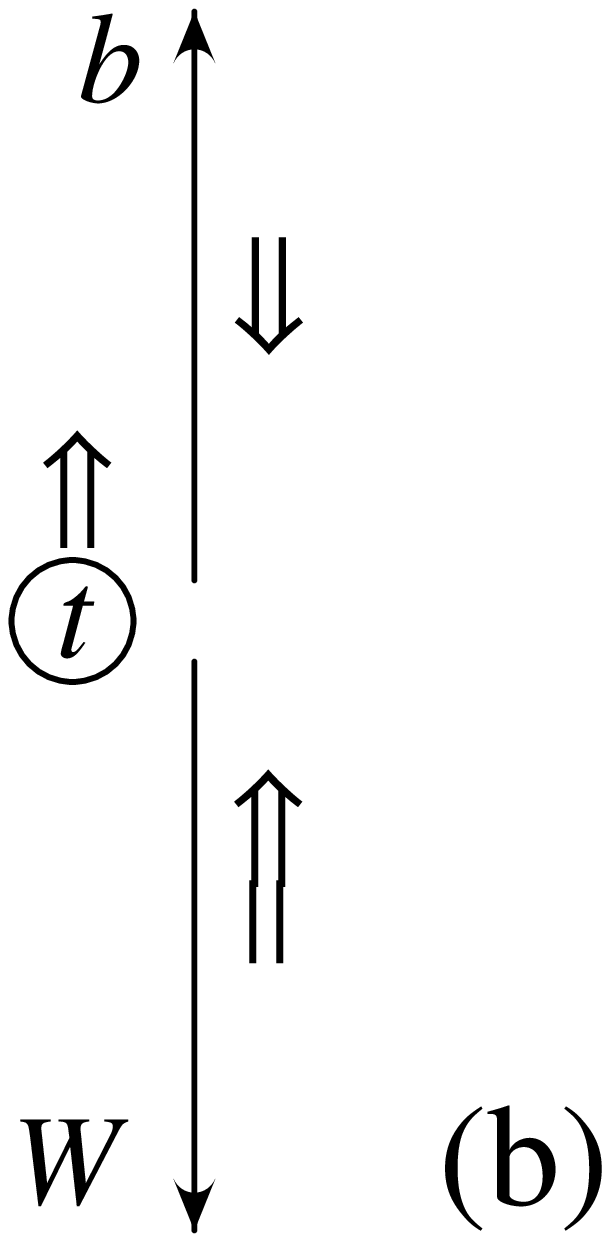}
\caption[]{Angular momentum conservation in the decay of
a polarized top quark.  Since the $b$ quark is effectively
massless and couples to a $W$, it is produced
with left-handed helicity.  (a)  $W$ emitted parallel to
top quark spin is longitudinal.  (b)  $W$ emitted antiparallel
to top quark spin is left-handed.}
\label{intuition}
\end{figure}

These correlations may be intuitively understood from elementary
angular momentum conservation arguments and the $V{-}A$ coupling
between the $W$ and quarks.  
Since the $b$ quark mass is much less than the energy
it receives from the decay,
the left-handed chirality of the
$tbW$ vertex translates into left-handed helicity for the $b$.
Suppose first that the $W$ boson is emitted along the top quark
spin axis, as in Fig.~\ref{intuition}a.  
Then, the spin of the $b$ points in the same direction
as the spin of the original $t$ and we must have 
zero spin projection for the $W$ boson ({\it i.e.}\ it must
be longitudinal).
On the other hand, when the $W$ is emitted in the
backwards direction (Fig.~\ref{intuition}b), the spin of the 
$b$ is opposite to the spin of the parent $t$.  
In this case, the $W$ must have left-handed
helicity in order to conserve angular momentum.

\section*{Acknowledgments}
I would like to thank Tim Stelzer and Stephen Parke
for the use of their figures in this talk.


\end{document}